\documentclass{appolb}
\usepackage{graphicx, slashed, listings, fancyhdr, amsmath, amsthm, amssymb, color, subfigure}
\begin{document}
\title{A dual order parameter from \\fundamentally colour charged matter%
\thanks{Presented at Light Cone 2012, Cracow, Poland}%
}
\author{M. Hopfer, M. Mitter, B.-J. Schaefer, R. Alkofer
\address{Institut f{\"u}r Physik, Karl-Franzens-Universit{\"a}t Graz\\ 
Universit{\"a}tsplatz 5, 8010 Graz, Austria}
\\
{}
}
\maketitle
\begin{abstract}
  The center phase transition of QCD and of fundamentally charged
  scalar QCD at non-vanishing temperature is investigated within a
  Dyson-Schwinger approach. The temperature dependence of the
  scalar/quark propagator is studied with generalized boundary
  conditions. A novel order parameter for the center phase transition
  is established which still exhibits a considerable dependence on the
  scalar/quark-gluon vertex.
\end{abstract}
  
\section{Motivation}

During the last years, intensive theoretical and experimental efforts
towards a deeper understanding of the confinement and chiral symmetry
breaking phenomena have been made.  In these proceedings we focus on
the QCD phase transition at non-vanishing temperature and
the related center symmetry breaking\footnote{Center symmetry is
  realized only in the limit of infinitely heavy quarks while in real
  QCD the symmetry is always explicitly broken,
  cf.~\cite{Polyakov:1978vu, Karsch:2003jg} and references therein.}  by investigating so-called
dual observables \cite{Gattringer:2006ci,
  Synatschke:2007bz,Fischer:2009wc,Fischer:2009gk,Fischer:2010fx,Braun:2009gm}. 
Dual observables can be related to
the spectrum of the Dirac operator and therefore comprehend
information on both, the chiral properties and the confinement of
quarks. They have first been introduced in lattice calculations
\cite{Gattringer:2006ci} but are also accessible via functional
methods \cite{Fischer:2009wc,Fischer:2009gk,Fischer:2010fx,Braun:2009gm}. Within a
Dyson-Schwinger approach \cite{Alkofer:2000wg} we study
the center phase transition in QCD and in fundamentally charged scalar
QCD.  For the solution of the Dyson-Schwinger equation (DSE) of the
scalar/quark propagator quenched lattice input for the gluon
propagator has been used \cite{Fischer:2010fx}. 
The main outcome of this investigation is
a novel order parameter for the center phase transition \cite{our_paper} 
which still exhibits a considerable dependence on the used
model for the scalar/quark-gluon vertex. This requires a more detailed
vertex analysis in future studies \cite{andreas_proceedings}.

\section{Center Phase Transition and Related Order Parameters}

The DSE for the quark propagator at finite temperature
reads\footnote{Here, for the quark-gluon vertex the renormalization
  factor $Z_{1F}=Z_2/\tilde Z_3$ has been used, while in Landau gauge
  the ghost-gluon vertex renormalization constant $\tilde Z_1=1$
  \cite{Taylor:1971ff}. Furthermore, the ghost renormalization
  constant $\tilde Z_3$ is omitted as it cancels by a corresponding
  factor in the quark-gluon vertex model. The remaining
  constant $Z_2$ is eliminated within a MOM scheme. $C_F$ is the fundamental
  Casimir invariant, $C_F=4/3$ for $N_c=3$.}
\begin{equation}
 S^{-1}(p) = Z_2 S_0^{-1}(p) - Z_2\,C_F\, g^2 T \sum_{\omega_k(\phi)} 
 \int\frac{d^3k}{(2\pi)^3}\, \gamma^\mu S(k) \Gamma^\nu(k,p;q) D^{\mu\nu}(q) 
\label{eq:quarkDSE}
\end{equation}
with the momenta $p=(\vec p,\omega_p)$. The generalized Matsubara
modes $\omega_p(\phi) = (2\pi T)(n_p+\phi/2\pi)$ depend on the general
boundary angle $\phi\in[0,2\pi)$.\footnote{A similar approach has been employed in 
\cite{Braun:2009gm}, where QCD with $\phi$-dependent boundary conditions for the quark fields has been 
referred to as $QCD_\phi$. This corresponds to QCD at imaginary chemical potential.} 
The usual fermionic boundary
conditions are obtained for $\phi=\pi$.  For the propagator we use the
conventional ansatz $S^{-1}(p) = i\gamma_4\omega_p C(p) +
i\slashed{\vec p} A(p) + B(p)$ and for the quark-gluon vertex the
temperature dependent model \cite{Fischer:2009gk}
\begin{equation}
 \begin{split}
   \Gamma^\nu(k,p;q) = \tilde Z_3 &
   \biggl(\delta^{4\nu}\gamma^4\frac{C(k)+C(p)}{2} +
   \delta^{j\nu}\gamma^j\frac{A(k)+A(p)}{2} \biggr) \\
   & \times\Biggl\{\frac{d_1}{d_2+q^2} +
   \frac{q^2}{q^2+\Lambda^2}\biggl( \frac{\beta_0\alpha(\mu)
     \ln\left[q^2/\Lambda^2+1\right]} {4\pi}\biggr)^{2\delta}\Biggr\}\ ,
 \end{split}
\label{eq:quarkgluonvertex}
\end{equation}
where $k$ and $p$ denote the in- and outgoing quark momenta and $q$
the gluon momentum.  The first part of the vertex
(\ref{eq:quarkgluonvertex}) consists of a Slavnov-Taylor motivated
ansatz, whereas the logarithmic tail together with the anomalous
dimension $2\delta=-18/44$ ensures a running coupling-like behaviour
of the vertex in the UV regime. The remaining model parameters $d_1$
and $d_2$ are purely phenomenological, where in the following we take
the values proposed in \cite{Fischer:2010fx}.  The system can now be
solved by using the gluon propagator as input from quenched lattice
results \cite{Fischer:2010fx}.

From the quark propagator \eqref{eq:quarkDSE}, various order
parameters can be extracted. The $\phi$-dependent quark condensate is
given by
\begin{equation}
 \langle\bar\psi\psi\rangle_\phi = Z_2 N_c T \sum_{\omega_p(\phi)}\int\frac{d^3p}{(2\pi)^3}\,
 tr_D\,S(\vec p,\omega_p(\phi))
\label{eq:quarkcondensate}
\end{equation}
where the case $\phi=\pi$ corresponds to the ordinary quark
condensate.  With a Fourier transform of the previous
condensate \eqref{eq:quarkcondensate} the dual quark condensate
\cite{Gattringer:2006ci} is obtained with respect to the
winding number $n$
\begin{equation}
 \Sigma_n = \int_0^{2\pi}\frac{d\phi}{2\pi}
 e^{-in\phi}\langle\bar\psi\psi\rangle_\phi\ .
\end{equation}
In the following we fix $n=1$. $\Sigma_1$ transforms like the
conventional Polyakov loop \cite{Polyakov:1978vu}
under center transformation and is therefore a suitable order
parameter for the center transition 
\cite{Gattringer:2006ci,Synatschke:2007bz,Fischer:2009wc,Fischer:2009gk,Fischer:2010fx}, 
cf. also \cite{Braun:2009gm}. 
Alternatively, the scalar quark
dressing function $B$, evaluated at vanishing external momenta, is
also sensitive to center symmetry breaking
\cite{Fischer:2009gk}.\footnote{A related order parameter is the 
dual quark mass parameter $\tilde M$ as proposed in \cite{Braun:2009gm}.}
However, if only the lowest
Matsubara mode is taken into account as suggested in
\cite{Fischer:2009gk} no satisfactory results could be obtained.
Hence, we propose the following order parameter
\vspace*{-0.2cm}
\begin{equation}
  \Sigma_Q =
  \int_0^{2\pi}\frac{d\phi}{2\pi}e^{-i\phi}\;\Sigma_{Q,\phi} \;,\quad
  \Sigma_{Q,\phi} =
  T\sum_{\omega_p(\phi)}\;\left[\frac{1}{4}tr_D\;S(\vec
    0,\omega_p(\phi))\right]^2 
\label{eq:orderparameter_quark}
\end{equation}
for the center phase transition. The novel condensate
$\Sigma_{Q,\phi}$ is periodic in the boundary conditions and
transforms like the conventional Polyakov loop under center
transformations \cite{our_paper}. In contrast to the quark condensate
$\langle\bar\psi\psi\rangle_\phi$ the new condensate is also finite
away from the chiral limit.

Before we substantiate our proposals with numerical results we proceed 
with the investigation of the center phase transition in scalar QCD.

\subsection*{Fundamentally Charged Scalar QCD}

Replacing the quarks by fundamental charged scalar fields
provides a simpler model system due to the absence of Dirac
structure. Neglecting the four-particle interactions the DSE for the
scalar propagator reads\footnote{For details on the derivation and the
  truncation scheme see e.g. \cite{Fister:2010yw} and references
  therein. We have omitted renormalization constants which cancel due
  to our scalar-gluon vertex model. Additionally, we apply a MOM
  scheme to eliminate $\hat Z_3$.}
\vspace*{-0.05cm}
\begin{equation}
 D_S^{-1}(p) = \hat Z_3(p^2+m^2) - \hat Z_3C_Fg^2 T \sum_{\omega_k}
 \hspace{-0.05cm}\int\hspace{-0.15cm}\frac{d^3k}{(2\pi)^3}(p+k)^\mu D_S(k) \Gamma_S^\nu(k,p;q) D_\gamma^{\mu\nu}(q) 
\end{equation}
with $D_S(\vec p,\omega_p) = Z_S(\vec p, \omega_p)/(\vec
p^2+\omega_p^2)$. For the scalar-gluon vertex we employ the model
\vspace*{-0.1cm}
\begin{equation}
 \begin{split}
 \Gamma_S^\nu(k,p;q) & = \tilde Z_3 \frac{D_S^{-1}(p^2)-D_S^{-1}(k^2)}{p^2-k^2} (p+k)^\nu\\ 
 & \times d_1\Biggl\{\frac{\Lambda^2}{\Lambda^2+q^2} + \frac{q^2}{q^2+\Lambda^2}\biggl(\frac{\beta_0\alpha(\mu)\ln\left[q^2/\Lambda^2+1\right]}{4\pi}\biggr)^{2\delta}\Biggr\}.
 \end{split}
\end{equation}
Here, the conventions for the momenta as well as the UV vertex
behaviour is equivalent to the one in Eq.\eqref{eq:quarkgluonvertex}.  
In addition,
we introduced a scalar propagator dependent function which is motivated
from Ward identities in scalar electrodynamics and represents a
generalized Ball-Chiu vertex \cite{Ball:1980ay}. The remaining part is
again purely phenomenological where a dimensionless modeling constant
$d_1=0.53$ is introduced which improves the description of the phase
transition. With the same lattice gluon propagator as in
\eqref{eq:quarkDSE}, the system can be solved\footnote{For this the program 
CrasyDSE \cite{Huber:2011xc} has been used.} 
and the order parameters can be extracted.

Similar to the previous QCD case the following object serves as an order
parameter for the center phase transition in scalar QCD
\begin{equation}
 \Sigma_S = \int_0^{2\pi}\frac{d\phi}{2\pi}e^{-i\phi}\;\Sigma_{S,\phi} \;,
 \quad \quad \Sigma_{S,\phi} = T \sum_{\omega_p(\phi)}D_S^2(\vec
 0,\omega_p(\phi))\ ,
 \label{eq:orderparameter_scalar}
\end{equation}
i.e., $\Sigma_{S,\phi}$ is periodic in $\phi$ and transforms like the conventional
Polyakov loop under center transformations \cite{our_paper}.

\section{Results}
\label{sec:results}

\begin{figure}
 \centering
%  \subfigure{\includegraphics[scale=0.5]{Plots/orderparameters.pdf}\label{fig:orderparameter_quarkA}}
 \subfigure{\includegraphics[scale=0.5]{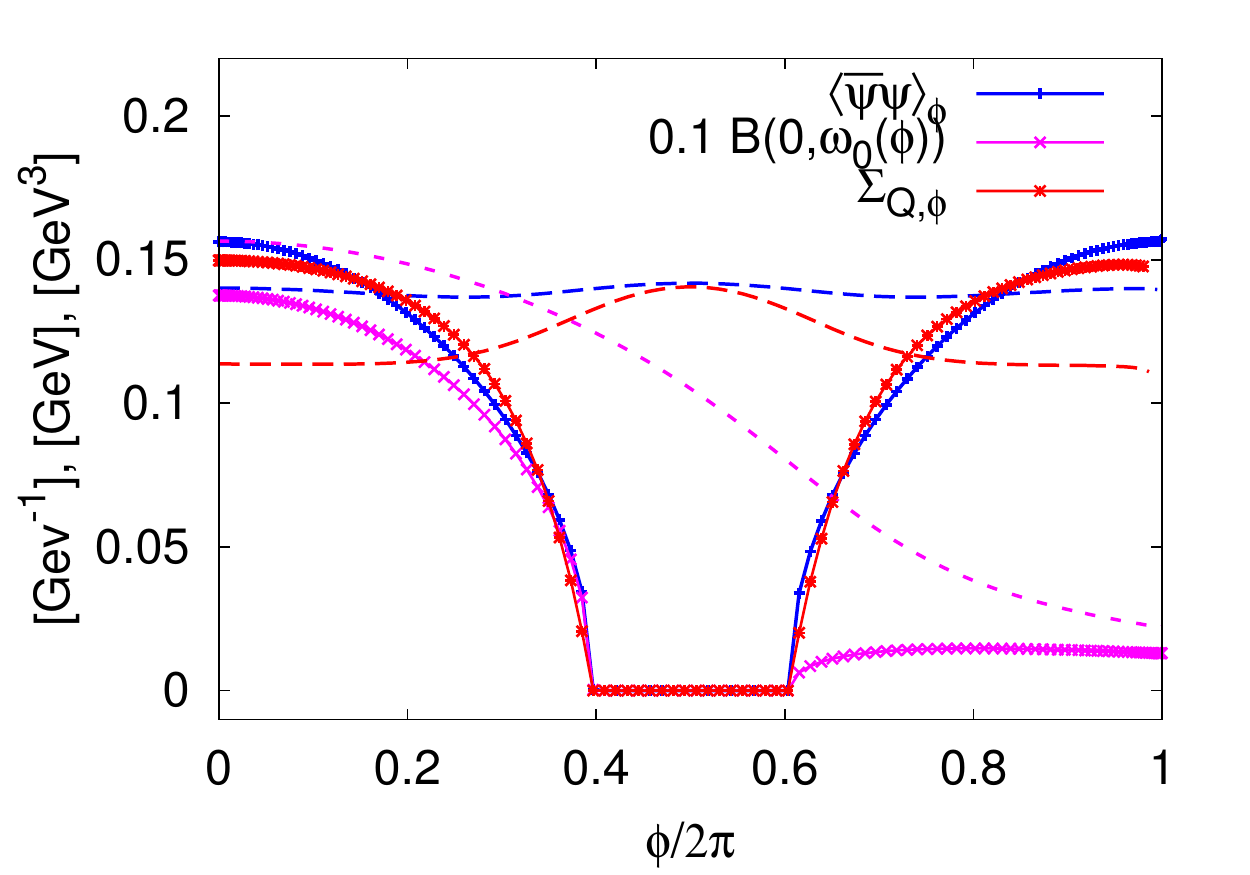}\label{fig:orderparameter_quarkA}}
 \hspace{-0.4cm}
%  \subfigure{\includegraphics[scale=0.5]{Plots/dualorderparameters.pdf}\label{fig:orderparameter_quarkB}}
 \subfigure{\includegraphics[scale=0.5]{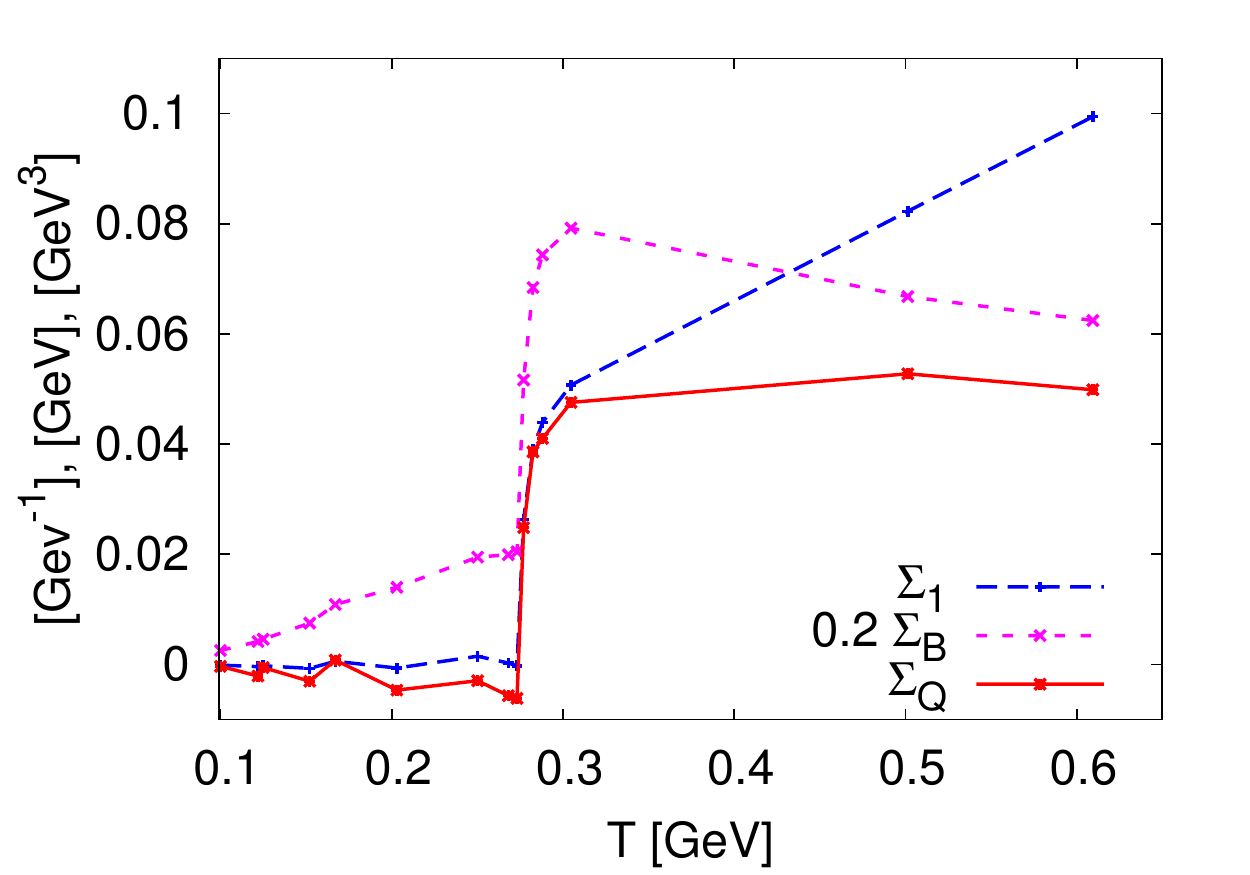}\label{fig:orderparameter_quarkB}}
 \caption{Left panel: The quark dressing function $B(\vec
   0,\omega_0(\phi))$, the quark condensates
   $\langle\bar\psi\psi\rangle_\phi$ and $\Sigma_{Q,\phi}$, as defined
   in Eq.\eqref{eq:orderparameter_quark}, as a function of the
   boundary angle for different temperatures in the chiral limit
   (dashed lines: $T=273$ MeV, solid lines: $T=283$ MeV). Right panel:
   The order parameters $\Sigma_{Q}$, $\Sigma_1$ and $\Sigma_B$ as
   defined in \cite{Fischer:2009gk} as a function of the temperature.}
\end{figure}

In order to confirm that $\Sigma_Q$,
Eq.\eqref{eq:orderparameter_quark}, and $\Sigma_S$,
Eq.\eqref{eq:orderparameter_scalar}, are well-defined order parameters
we consider both QCD and scalar QCD at finite temperatures. In 
Fig.\ref{fig:orderparameter_quarkA} the $\phi$-dependence of
$\Sigma_{Q,\phi}$ and the quark condensate
$\langle\bar\psi\psi\rangle_\phi$ are shown. While both quantities are
periodic in $[0,2\pi)$ and symmetric in $\phi\rightarrow-\phi$ the
scalar quark dressing function $B(p)$, evaluated at vanishing momenta
and lowest Matsubara mode, is not periodic. Above the critical
temperature $T_c=277$ MeV all three quantities melt and, in the chiral
limit, a plateau is formed. Below $T_c$ the quark condensate is 
nearly constant while $\Sigma_{Q,\phi}$ is slightly enhanced around
$\phi=\pi$. This yields a non-zero value of $\Sigma_Q$ below $T_c$ as
compared to the dual condensate $\Sigma_1$ which almost vanishes in
this region as shown in Fig.\ref{fig:orderparameter_quarkB}. Note
that the temperature behaviour of $\Sigma_Q$ below the phase
transition can be further tuned by varying the vertex model
parameters. In contrast, the order parameter $\Sigma_B$, as defined in
\cite{Fischer:2009gk}, considerably deviates from zero already below
the phase transition, whereas $\Sigma_1$ and $\Sigma_Q$ stay close to
zero. In order to see the expected behaviour of the order parameters
the incorporation of all Matsubara modes is mandatory.

\begin{figure}[htb]
  \centering
%   \subfigure{\includegraphics[scale=0.5]{Plots/FMM_su3_dual_phi.pdf}\label{fig:orderparameter_scalarA}}
  \subfigure{\includegraphics[scale=0.5]{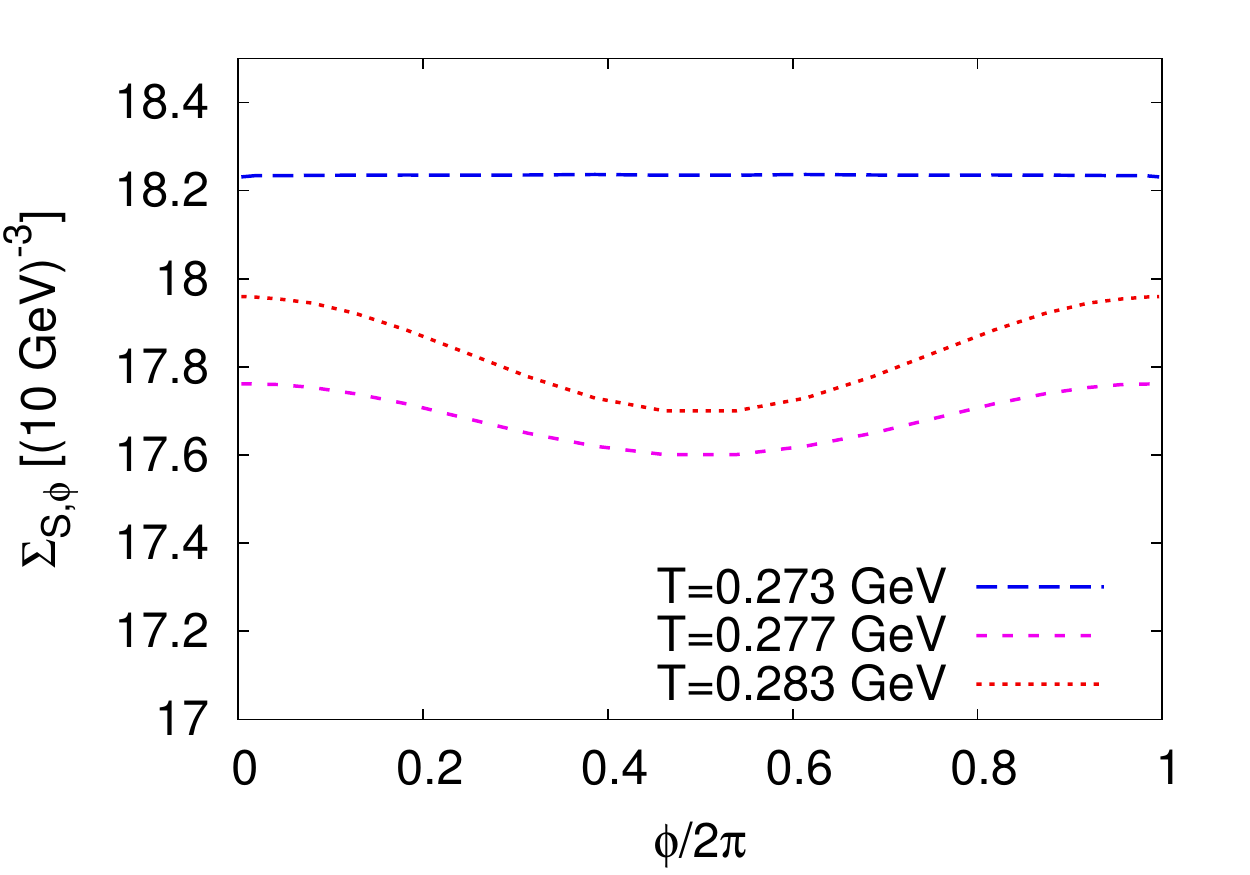}\label{fig:orderparameter_scalarA}}
  \hspace{-0.4cm}
%   \subfigure{\includegraphics[scale=0.5]{Plots/FMM_su3_dual_vert_T.pdf}\label{fig:orderparameter_scalarB}}
  \subfigure{\includegraphics[scale=0.5]{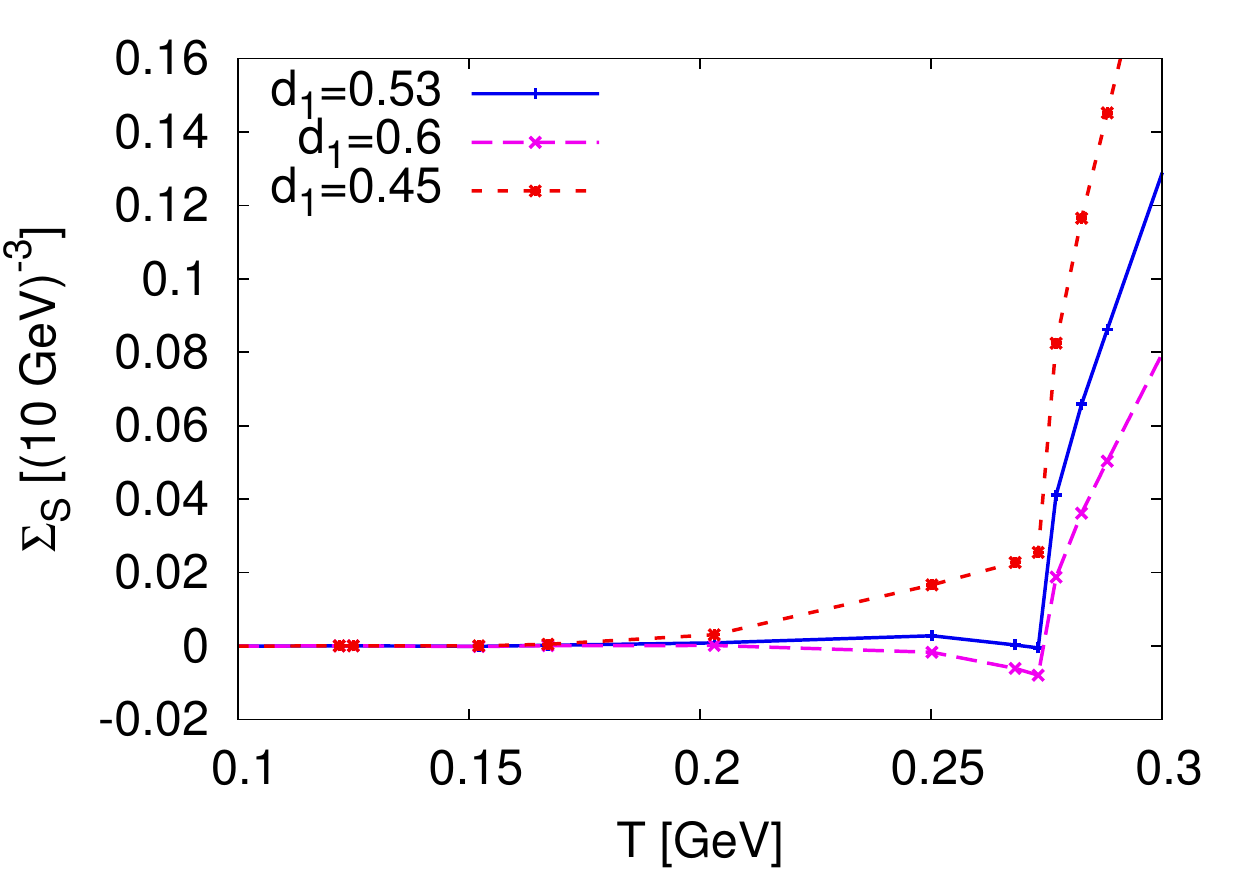}\label{fig:orderparameter_scalarB}}
  \caption{Left panel: $\Sigma_{S,\phi}$ as a function of the boundary
    angle for three different temperatures and $m=1.5$ GeV. Right
    panel: The order parameter $\Sigma_S$ as a function of the
    temperature for three different vertex parameter values.}
\end{figure}

For scalar QCD we obtain similar results as is demonstrated in
Fig.\ref{fig:orderparameter_scalarA} where the $\phi$-dependence of
the condensate $\Sigma_{S,\phi}$ is shown.  Below $T_c$ the condensate
is $\phi$-independent and therefore $\Sigma_S$ vanishes while above $T_c$
it is suppressed around $\phi=\pi$ and hence $\Sigma_S$ is
non-vanishing. This is displayed in
Fig.\ref{fig:orderparameter_scalarB} where a distinct phase transition
is observed. However, the results strongly dependent on the used
scalar-gluon vertex model.

\section{Conclusions}

Within the Dyson-Schwinger formalism we investigated the center phase
transition of QCD and fundamentally charged scalar QCD. A novel order
parameter was introduced which exhibits an improved behaviour below
$T_c$ compared to the order parameter $\Sigma_B$ proposed in
\cite{Fischer:2009gk}. Motivated by the
strong model dependence on the used
vertex further studies of the corresponding scalar/quark-gluon vertex
are on-going, see e.g. \cite{andreas_proceedings}.

\section*{Acknowledgements}
We thank C.S. Fischer, L. Fister, J. Luecker, A. Maas, P. Maris and
J.M. Pawlowski for valuable discussions. This work was supported by
the Research Core Area 'Modeling and Simulation' at the University of
Graz. MH and MM were funded by the Austrian Science Fund, FWF,
through the Doctoral Program on Hadrons in Vacuum, Nuclei and Stars
(FWF DK W1203-N16) and BJS by the FWF under grant number P24780-N27.
\goodbreak

% \section*{References}

%\bibliographystyle{bibstyle}
%\bibliography{MyBibTexFile}

%uncomment the following lines to place a figure
%\begin{figure}[htb]
%\centerline{%
%\includegraphics[width=12.5cm]{Fig1}}
%\caption{Plot of ...}
%\label{Fig:F2H}
%\end{figure}

\end{document}